\documentclass[
aps,
prl,
twocolumn,
amsfonts,
amssymb,
amsmath,
superscriptaddress,
showpacs,
]{revtex4-1}
\usepackage{everypage}
\usepackage{amsmath}
\usepackage{amssymb}
\usepackage%
{graphicx}
\usepackage{extdash}
\usepackage{epsfig}
\usepackage{textcomp}
\usepackage{letltxmacro}
\usepackage{datetime}

\DeclareFontFamily{U}{euc}{}
\DeclareFontShape{U}{euc}{m}{n}{<-6>eurm5<6-8>eurm7<8->eurm10}{}%
\DeclareSymbolFont{AMSc}{U}{euc}{m}{n} 
\DeclareMathSymbol{\umu}{\mathord}{AMSc}{"16} 

\renewcommand{\vec}[1]{\boldsymbol{#1}}
\newcommand{\ensuretext}[1]{\ensuremath{\text{#1}}}

\newcommand{\unit}[1]{\ensuretext{\textrm{\,}}\ensuremath{\mathrm{#1}}}
\newcommand{\eV}{\textit{e}\mathrm{V}}

\newcommand{\keV}{\mathrm{k}\eV}

\newcommand{\Mum}{\ensuremath{\umu}\ensuremath{\mathrm{m}}}
\newcommand{\mum}{\textrm{\,\ensuremath{\mathrm{\Mum}}}}
\newcommand{\BoundE}{\Gamma}
\newcommand{\BoundLowE}{\gamma}

\begin{document}


\title{Nanoscale femtosecond imaging of transient hot solid density plasmas with elemental and charge state sensitivity using resonant coherent diffraction}

\author{T.~Kluge}
\email[]{t.kluge@hzdr.de}
\homepage[]{http://hzdr.de/crp}
\affiliation{Helmholtz-Zentrum Dresden-Rossendorf, Germany}

\author{M.~Bussmann}
\affiliation{Helmholtz-Zentrum Dresden-Rossendorf, Germany}

\author{H.-K. Chung}
\affiliation{International Atomic Energy Agency, Austria}

\author{C.~Gutt}
\affiliation{Universit\"at Siegen, Germany}

\author{L.\,G.~Huang}
\email[]{lingen.huang@hzdr.de}
\affiliation{Helmholtz-Zentrum Dresden-Rossendorf, Germany}

\author{M.~Zacharias}
\affiliation{Helmholtz-Zentrum Dresden-Rossendorf, Germany}

\author{U.~Schramm}
\affiliation{Helmholtz-Zentrum Dresden-Rossendorf, Germany}
\affiliation{TU Dresden, Germany}

\author{T.\,E.~Cowan}
\affiliation{Helmholtz-Zentrum Dresden-Rossendorf, Germany}
\affiliation{TU Dresden, Germany}

\date{\today}

\begin{abstract}
Here we propose to exploit the low energy bandwidth, small wavelength and penetration power of ultrashort pulses from XFELs for resonant Small Angle Scattering (SAXS) on plasma structures in laser excited plasmas. 
Small angle scattering allows to detect nanoscale density fluctuations in forward scattering direction. 
Typically, the SAXS signal from laser excited plasmas is expected to be dominated by the free electron distribution. 
We propose that the ionic scattering signal becomes visible when the X-ray energy is in resonance with an electron transition between two bound states (Resonant coherent X-ray diffraction, RCXD). 
In this case the scattering cross-section dramatically increases so that the signal of X-ray scattering from ions silhouettes against the free electron scattering background which allows to measure the opacity and derived quantities with high spatial and temporal resolution, being fundamentally limited only by the X-ray wavelength and timing. 
Deriving quantities such as ion spatial distribution, charge state distribution and plasma temperature with such high spatial and temporal resolution will make a vast number of processes in shortpulse laser-solid interaction accessible for direct experimental observation e.g. hole-boring and shock propagation, filamentation and instability dynamics, electron transport, heating and ultrafast ionization dynamics. \\
\end{abstract}


\maketitle
\section{Introduction}
One of the essential elements in the generation of transient hot solid-density plasmas by ultra-high intensity (UHI) lasers is the relativistic electron generation~\cite{Mishra2009,micheau2010generation,Sentoku2002} and transport dynamics~\cite{Sentoku2002,Okabayashi2013,Chawla2013,Zhuo2013,Leblanc2014}. 
Near the laser focus, the current density of the relativistic electrons can exceed $10^{13}\unit{A/cm}^2$~\cite{Sentoku2002}. 
This is intimately intertwined with the ionization dynamics and the complex evolution of the bulk return currents~\cite{Sentoku2007,LGHuang}. 
These are important due to the strong magnetic and electric fields generated at these current densities, up to $10^5~\unit{T}$ and $10^{14}\unit{V/m}$, as well as the rapid temporal and spatial evolution of the bulk temperature, ionization state, and hence resistivity by virtue of the electron-ion collision frequency, and anomalous resistivity from strong fields~\cite{Sentoku2007,LGHuang,MacLellan2013,Chawla2013,Zhuo2013,Leblanc2014}. \\
At present, a predictive understanding of high-intensity laser-matter interactions is severely hampered by the lack of self-consistent models for the ionization and recombination dynamics, coupled with the complex electron transport and collisions, and our inability to unravel this complexity with available experimental techniques in laser-only experiments. 
For example, effective ionization rates, scattering cross-sections and K$\alpha$ and other self-radiation spectra are not precisely known in a non-thermal non-static dense plasma~\cite{Mishra2013}. 
Additionally, numeric simulations are limited e.g. due to the short time-step necessary for kinetic simulation of fast plasma oscillations, which can even reach classical electron orbit rotation times in an ion~\cite{MGarten}. 
Due to the large particle numbers in solid density plasmas the use of a full quantum mechanical ionization description will surely remain a dream for the next decades.

With the availability of fourth-generation X-ray free electron laser (FEL) sources and their instrumentation with ultra-intense short-pulse lasers many novel experiments can be envisioned that will e.g. enhance the diagnostic capabilities for femtosecond laser-solid interactions. 
Due to short duration of XFEL pulses and their penetration power through solids, coherent scattering allows probing of processes on femtosecond \emph{and} sub-micron scale inside the solid plasma for the first time. 
Measuring the plasma opacity with high spatial and temporal resolution would allow for the first time to directly obtain experimentally the ion distribution during short-pulse ultra-intense laser interaction with solids and compare directly to simulations and ionization models used therein, or to study the plasma and hot electron dynamics through their influence on ionization states. \\
We will show how the low energy bandwidth, small wavelength and penetration power of ultrashort pulses from XFELs can be used for Small-Angle Scattering (SAXS) on plasma structures~\cite{Kluge2014} to not only scatter on free electrons but also to resonantly scatter at selected ions (resonant coherent X-ray diffraction, RCXD). 
This allows to obtain information with high temporal resolution about the spatial structure of the distribution of electrons and ions (e.g. filamentation, hole-boring, hot electron divergence and surface structures) as well as the energy distribution of free electrons (i.e. plasma temperature)\cite{Marjoribanks1992,Gregori2005,Chen2009} with few nanometer resolution at the same time, as it is not possible with any other available technique. 
It would allow unique tests of models e.g. for non-equilibrium ionization dynamics, heating, resistivity, electron transport or laser absorption. 
Pump-probe or pulse split-and-delay experiments would enable the dynamic measurement of the quantities, e.g. shock velocity, ion front and hole-boring velocity, or correlation spectroscopy. \\
Resonant Elastic X-ray Diffraction or Anomalous Small-Angle X-ray Scattering~\cite{RCXDI,Scherz2008,joly2012resonant} is a standard X-ray technique employing bound-free resonances that typically involve outer electron shells influenced by the surrounding structure of the atom to infer information on chemical or biological properties or physical structure (e.g of the lattice, spin distribution or magnetic domains) of a sample or molecule. 
In the present paper we discuss the possibilities that open up when transferring the methodology to bound-bound resonances in laser-generated plasmas in order to be sensitive to the ionic scattering signal and much less susceptible to many-body phenomena and plasma screening. \\

\section{Scattering in warm plasmas}
\begin{figure}
	\includegraphics[width=8cm]{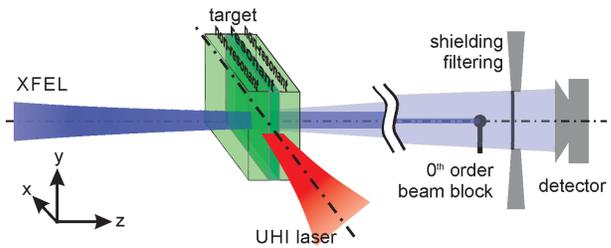}
	\caption{Schematic setup of a RCXD experiment with a buried resonant layer target. Here the UHI is oriented normal to the target foil front surface and the XFEL penetrates the buried layer. The scattered photons are recorded far behind the target with a 2D areal detector after shielding and spectral filtering. }
	\label{fig:setup}
\end{figure}

The ionic scattering signal becomes visible against the non-resonant free electron scattering background when the X-ray energy is in resonance with an electron transition between two bound states. 
In this case the scattering cross-section dramatically increases due to the optical correction so that the signal of resonant X-ray scattering from ions silhouettes against the free electron scattering background despite the lower ion density. 
A schematic setup of a laser-plasma RCXD experiment is sketched in Fig.~\ref{fig:setup} with the UHI laser and XFEL intersecting at the solid foil at approximately $90^\circ$ to each other. 
Important parameters used in the following are summarized in Tab.~\ref{tab:parameters}. \\
\begin{table}
  \begin{minipage}[b]{0.48\textwidth} 
	 \parfillskip=0pt
   \begin{minipage}[!t]{0.49\textwidth} 
	  \tiny
		\squeezetable
		\begin{tabular}{ll}
		&\\
				\textbf{UHI laser} & \\
		  \hline
		  laser strength     & $a_0=6.8$\\
  		duration & $\tau=40 \unit{fs}$ (FWHM)\\
	  	wavelength & $\lambda=0.8\mum$\\
	  	wavelength & $\lambda=0.8\mum$\\
	    \textbf{XFEL}      & \\
		  \hline
			irradiated area & $A=8\mum\times8\mum$\\
		  fluence   & $10^{10}$ photons/A\\
			$E_{XFEL}$& $8.290\keV$\\
			energy spread&$20\eV$\\
  		arrival time & $t_{arr}=80\unit{fs}$ past UHI\\
	  	duration & $\ll$ typ. ion motion\\
    \end{tabular}	
	\end{minipage}
	\hfill
  \begin{minipage}{0.49\textwidth} 
	  \tiny
		\squeezetable
	  \begin{tabular}{ll}
			\textbf{Plasma}&\\
			\hline
			sim. box size&$7.2 \times 21.6\mum^2$\\
			cells/\mum & 250\\
			material & copper\\
			foil thickness&$d=1.6\mum$ in x-dir\\
			preplasma&exp., scale $0.08\mum$\\
			particles per cell&10 ions\\
			(if fully ionized)&290 e- \\
    \end{tabular}	\vspace{0.9cm}
	 \end{minipage}
  \end{minipage}
	\caption{Summary of important parameters used in this paper.}
	\label{tab:parameters}
\end{table}

\begin{figure}
	\includegraphics[width=8cm]{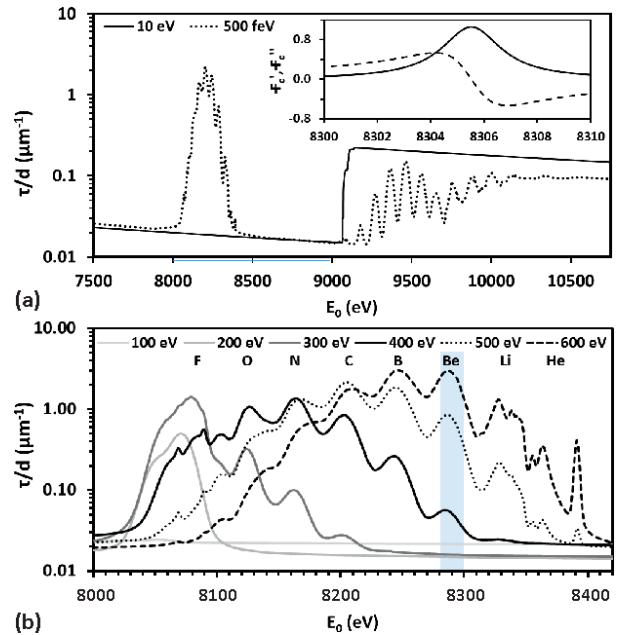}
	\caption{Opacity of a copper plasma slab at different temperatures calculated by SCFLY~\cite{flychk,*scfly}. The inset in (a) shows exemplary the optical corrections for the specific configuration $c=1s^22s^2$ at its $K\alpha_1$ transition energy. The highlighted energy range indicates the XFEL energy and bandwidth used for the simulations in Fig.~\ref{fig:results}. (Note, the cold $K$-edge and $K\alpha$-resonances obtained from SCFLY exhibit a small shift compared to the NIST values~\cite{NIST_XCOM}, which however is not relevant for the qualitative morphology.) }
	\label{fig:opacity}
\end{figure}

For a cold target the X-ray interaction is dominated by photoelectric absorption, see Fig.~\ref{fig:opacity}a for the example of copper around the $K$-edge. 
Once the temperature is sufficiently high and lower shells are not fully filled anymore due to excitation or ionization, channels for bound-bound transitions open up at photon energies just lower than the respective cold ionization edge, seen in Fig.~\ref{fig:opacity}a for a temperature of $500\unit{\eV}$ for $K\alpha$ transitions in copper. 
When the XFEL energy $E_0$ matches that of a bound-bound transition of an ion in a certain electronic configuration $c$, it will excite electron transitions between the two bound states accompanied by emission of X-rays. 
This can be treated as resonant scattering of the XFEL and be described by the complex valued change $F_c^{res}=F'_c(E_0)+iF''_c(E_0)$ of the ionic scattering form factor $F_0=F_{\BoundE}\cong\BoundE$~\cite{joly2012resonant} (inset in Fig.~\ref{fig:opacity}a). 
Here, $\BoundE=Z-Q$ is the number of bound electrons of an ion with $Z$ protons and total charge $Q$, the form factor $f_{0}$ of an electron is equal to unity since we consider only the limit of small scattering angles and we neglect the internal electronic structure of the ions. \\
For the more realistic case of finite XFEL FWHM  bandwidth $\Delta E$ and a mixture of ions with different configurations $F^\mathrm{res}_c$ as found in warm plasmas, the average optical corrections to the form factor of ions with $\BoundE$ bound electrons are given by the sum over all configurations of a charge state weighted by their abundance $n_{c}(\vec{r})$ and averaging over the XFEL energy profile,
$\bar{F}^\mathrm{res}_{\BoundE}\equiv\frac{1}{n_{\BoundE}}\sum_\mathrm{c~in~\BoundE} \left\langle n_{c} F^\mathrm{res}_c(E)\right\rangle_E$. 
The imaginary part of the optical correction caused by a single resonance can be derived from the product of its transition probability and the relative lower level population. 
The sum over all resonances in all ions is proportional to the opacity  $\tau(E_0)$, $\tau(E_0)=-\frac{4\pi c r_0 \hbar d}{E_0}\sum_{\BoundE=1}^{Z} n_{\BoundE} F''_\BoundE$~\cite{*[{}] [{, page 284}] Als-Nielson2011} ($d$ is the foil depth and $r_0$ is the classical electron radius). 
Using Kramer-Kronig's relation, the real part of the optical correction can be obtained. 

\begin{figure}
	\includegraphics[width=8cm]{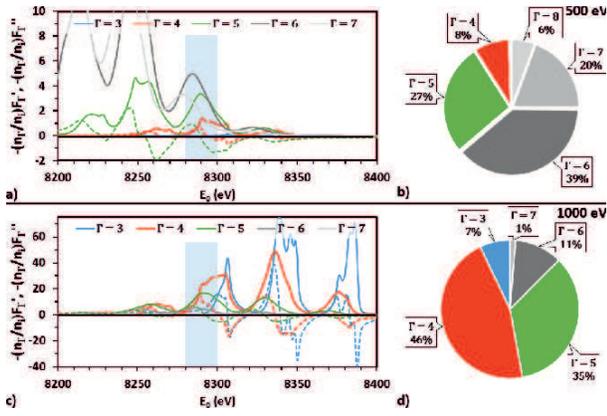}
	\caption{Contribution of different ion species to the plasma opacity $\tau\propto n_{\BoundE}F''_\BoundE$ (solid lines) and $n_{\BoundE}F''_\BoundE$ (dashed lines) for Copper ions with $\BoundE=3$ to $7$ bound electrons and plasma temperature $500\unit{\eV}$ (a,b) and $1000\unit{\eV}$ (c,d) from SCFLY. (b) and (d) show the relative magnitudes of the contribution to the opacity for different ion charges, averaged over $8280\unit{\eV}-8300\unit{\eV}$. }
	\label{fig:configuration_Q_weighted}
\end{figure}

\begin{figure}
	\includegraphics[width=8cm]{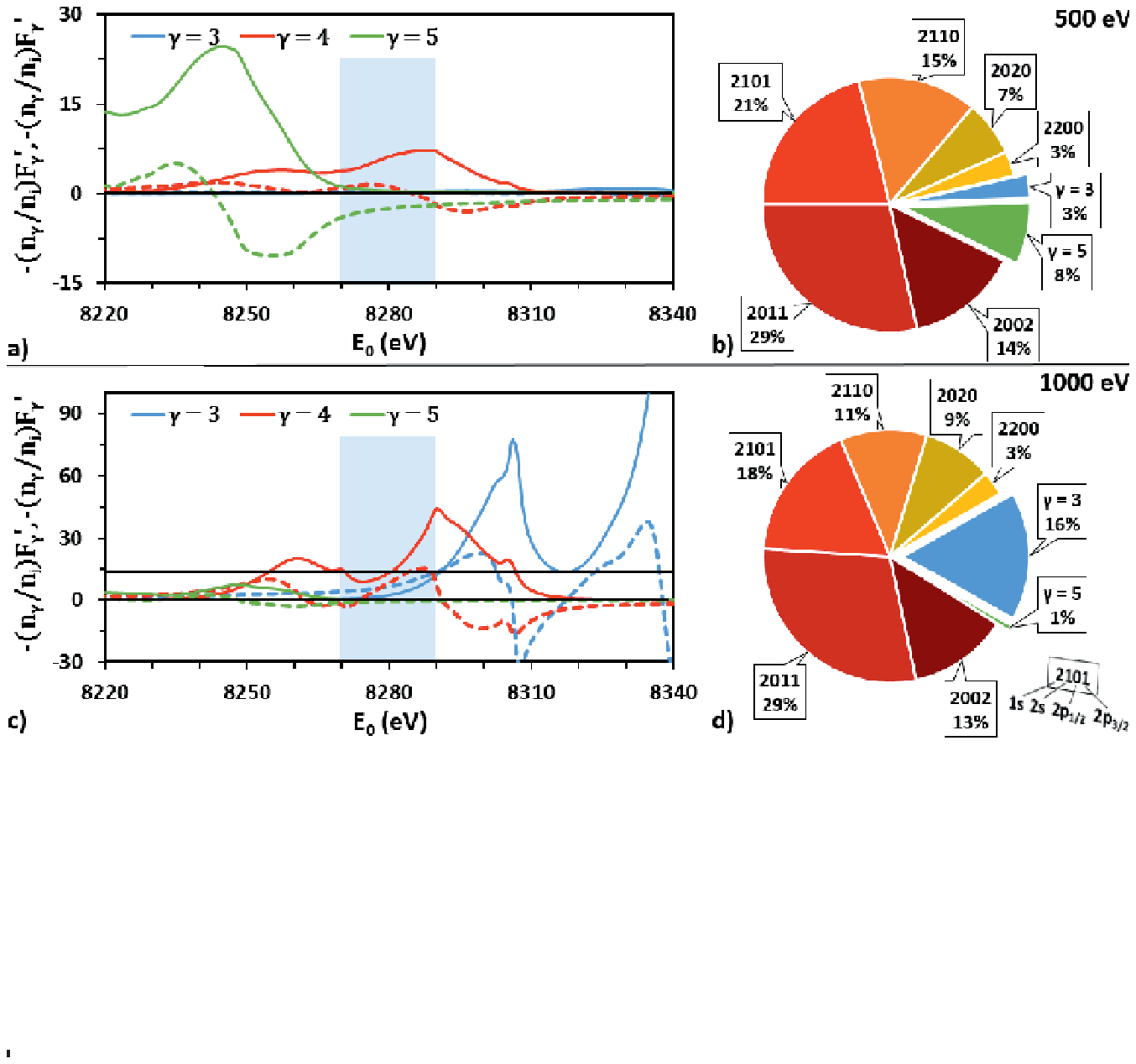}
	\caption{Contribution of different ion configurations of the K and L shells to the plasma opacity $\tau$ for Copper ions at plasma temperature $500\unit{eV}$ (a,b) and $1000\unit{eV}$ (c,d) from SCFLY calculations. Ions with $\BoundLowE=4$ (red, orange) dominate regardless of plasma temperature. The highest specificity up to $1\unit{\keV}$ plasma temperature is obtained at $E_0=8275\unit{\eV}$. Then, at $500\unit{\eV}$, ions with $\BoundLowE=4$ contribute almost $100\%$, at $1000\unit{eV}$ still $\approx 90\%$}
	\label{fig:configuration_inner_weighted}
\end{figure}

Figure~\ref{fig:opacity}b shows the opacity for the example of Copper for various plasma temperatures calculated by SCFLY~\cite{flychk,*scfly} with distinct maxima visible. 
The typical naming scheme is following the resonant energy of ground state ions with $\BoundE$ electrons, e.g. the He-, Li- and Be-like peaks for $\BoundE=2,3,4$ respectively. 
Fig.~\ref{fig:configuration_Q_weighted} shows the contribution of specific ion charges to the total opacity for two plasma temperatures. 
For cold plasma, for example at the opacity Be-like peak around $8290\unit{\eV}$, ions with 4 remaining electrons are responsible for less than $10\%$ of the opacity only. 
Due to the increase of abundance of highly charged ions at the expense of less charged ions, from $500\unit{eV}$ to $1000\unit{eV}$ temperature the contribution of ions with 5 electrons to the Be-like opacity peak decreases while that of ions with $\BoundE=4$ and $\BoundE=3$ increases. 
Yet, at no temperature does a single ion species dominate. 
In warm plasmas generally ions with different numbers of bound electrons contribute significantly to the total opacity at most XFEL energies. \\
The reason is that spectator electrons in those shells not significantly screening the nuclear potential as seen by electrons participating in a transition do not influence its resonant energy, e.g. excited ions with specific $\BoundE$ can have similar resonant energies as a ground state ion with electron number $\BoundE-1$.
To give a specific example, for K$\alpha$ transitions an ion with $\BoundE=5$ remaining electrons being in the excited state $1s^22s^23p^1$ with one spectator electron in the M-shell has generally similar resonant energies for transitions between the 1s and 2p$_{1/2}$ and 2p$_{3/2}$ levels as does an ion with $\BoundE=4$ in the ground state $1s^22s^2$. \\ 
Therefore, in order to identify resonances, the number $\BoundLowE$ of electrons in the lower shells that do influence the transition energy of the respective resonance is a better parameter than $\BoundE$ (in the above example the K and L shells).  
In analogy to $\bar{F}^\mathrm{res}_\BoundE$ we define 
\begin{equation}
  \bar{F}^\mathrm{res}_{\BoundLowE}\equiv\frac{1}{n_{\BoundLowE}}\sum_\mathrm{c~in~\BoundLowE} \left\langle n_{c} F^\mathrm{res}_c(E)\right\rangle_E. 
	\label{eqn:FQ}
\end{equation}
extending the sum over all configurations with the same respective $\BoundLowE$ regardless of the configuration of outer shells. \\
Resonant scattering at a specific $E_0$ can generally be attributed primarily to a single specific number of electrons in lower shells, justifying the labels of the opacity peaks given in Fig.~\ref{fig:opacity}b retrospectively with a slightly other meaning. 
E.g. from Fig.~\ref{fig:configuration_inner_weighted} it can be seen that for the example of the Be-like peak around $8290\unit{\eV}$ ions with $\BoundLowE=4$ dominate the resonant scattering largely independent of the plasma temperature. 

\section{Resonant Small Angle Scattering from ultrashort laser irradiated solids}
The flux $I(k_x,k_y)$ of scattered X-rays from scattering off electrons and ion resonances is given in first Born approximation by
\begin{align}
\label{eqn:scatter}
	I(k_x,k_y)&=\left.\tilde{I}(\vec{k})\right|_{k_z=0} \mathrm{where } \tilde{I}(\vec{k})\propto H(\vec{k})H(\vec{k})^*\\
  \nonumber H(\vec{r})&= h^\mathrm{free}(\vec{r})+h^\mathrm{bound}(\vec{r})+h^\mathrm{res}(\vec{r})\\
	\nonumber &\equiv n^\mathrm{free}_e(\vec{r})f_0+\sum_{\BoundE=1}^{Z}{n_{i,\BoundE}(\vec{r}) F_\BoundE}+\sum_{\BoundLowE=1}^{N-1}{n_{i,\BoundLowE}(\vec{r})F^\mathrm{res}_\BoundLowE(\vec{r})}
\end{align}
where $H(\vec{k})$ is the Fourier transform of $H(\vec{r})$, assuming a quasi-static plasma during the XFEL pulse, XFEL pulse propagation along the $z$-axis and $N$ is the maximum occupation number up to the transition's upper level. 
Near bound-bound resonances $\left|F^{res}_\BoundLowE(\vec{r})\right|$ increases rapidly and can reach values well exceeding $100$, then dominating over the non-resonant contributions. 
This is the reason for strong ion contrast with elemental and inner shell charge state specificity in the scattering signal even at lower ion density than that of free electrons. 
Additionally the non-resonant free and bound electrons contribute dominantly to signal in other $\vec{k}$ regions than the resonant signal as the target keeps net neutrality on sufficiently large scales, and hence can generally be easily discriminated from the latter. 

A schematic setup of a laser-plasma RCXD experiment was shown in Fig.~\ref{fig:setup} with the UHI laser and XFEL intersecting at the solid foil at approximately $90^\circ$ to each other. 
In such a setup the XFEL penetrates the target foil in z-direction and is scattered at structures that are in the $x-y-$plane, e.g. filaments, shock fronts hole-boring compressions, plasma oscillations and others. 
The scattering pattern can be recorded behind the target with a 2D area photon detector. 
A simple experimental realization of the setup could be a flat foil with the XFEL beam intersecting the optical laser slightly less than normal, penetrating the foil at a shallow angle. 
We note that in principle the presented RCXD method could also be used in a grazing incidence SAXS (GI-SAXS) geometry with X-ray penetration and scattering limited to the foil surface. 
The scattering pattern contains the full mode pattern of all electron correlations as well as that of resonating ions. 
The real space distribution $H(\vec{r})$ can eventually be reconstructed using phase retrieval algorithms or by employing more sophisticated holographic methods~\cite{Fienup1982,McNulty1992,Stadler2008,Mancuso2010}.

The scattering pattern $I(\vec{k})$ does not only contain information on the structure of spatial ion distribution but moreover information on distribution of charge states themself and plasma temperature. 
This information is encoded in the relative magnitude of real and imaginary part of the optical corrections, and it manifests itself in an asymmetry of the scattering pattern. 
If the XFEL energy $E_0$ is tuned to match a resonance in ions with a specific $\BoundLowE$, also ions with more or less electrons in the lower shells may contribute as the resonances have a non-vanishing width (Fig.~\ref{fig:configuration_inner_weighted}). 
The contribution of such neighboring resonances manifests itself not only through the imaginary part of the optical correction, but also through the real part. 
Specifically, the real part of the optical correction at the center of a resonance vanishes, so any non-vanishing real part is due to neighboring resonances. 
Negative values of the optical correction real part are due to resonances at higher energy (smaller $\BoundLowE$) while positive values are due to resonances at lower energy (larger $\BoundLowE$)
As the spatial distribution of ions with different values of $\BoundLowE$ is generally different, also the real part of the optical correction will be distributed differently than the imaginary part, breaking the point-symmetry $I(\vec{k})=I(\vec{-k})$ of the scattering signal. 
This effect can be utilized by phase retrieval methods to retrieve the imaginary and real parts separately and hence to infer the relative amount of ions with different charge state than that selected by $E_0$. 
This is especially interesting as different ionization models employed in state-of-the-art simulation codes predict different distributions, thus allowing for a direct experimental validation. 

Furthermore, the ratio of real part to imaginary part of the optical correction, as well as relative amplitudes in a two-color XFEL experiment, can be used as a thermometer for the plasma temperature.
The local electron temperature, or more generally the electron energy distribution, and its temporal evolution, determines the population of individual electron configurations and therefore the spectral shape of the optical corrections~\cite{Marjoribanks1992,Gregori2005,Chen2009}. 
As the number of electron holes in lower shells increases with plasma temperature, the real part of the optical correction shifts from positive for small temperatures to negative for higher temperatures. 
For a thermalized system this effect can be seen in Fig.~\ref{fig:configuration_inner_weighted} for the example of copper ions, as simulated by SCFLY. 
The measurement of real and imaginary part of the optical correction can hence be used -- in conjunction with an appropriate dynamic model -- to determine the plasma temperature. 

So far, we discussed the change in the optical correction real and imaginary part which has to be derived from the asymmetry induced into the scattering data via phase retrieval. 
However, using simultaneously multiple XFEL pulses with varying wavelength the exact spectral shape of the opacity curve and hence charge state distribution and temperature can be obtained directly experimentally. \\

\begin{figure*}
	\includegraphics[width=\textwidth]{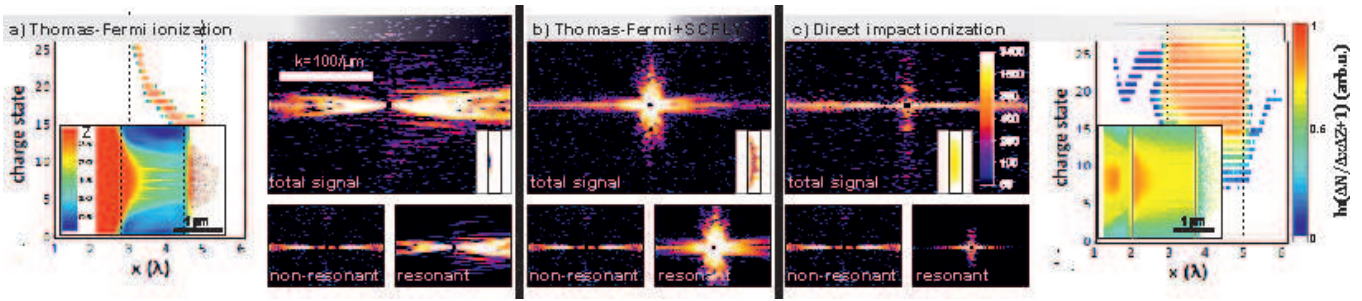}
	\caption{Simulated longitudinal charge state distribution along the laser axis (logarithmic scale) and 2D average charge state from PIC simulations employing TF (a) and DI ionization (c) (target position is marked by dashed lines) together with the simulated scattering images (number of scattered photons per detector pixel) on a detector $1\unit{m}$ behind the target with pixel size $15\unit{\mum}$. Scattering signal for TF ionization with the ion distribution being constructed  based on the plasma electron temperature by SCFLY during post-processing of PIC simulation is shown in (b). The signals as expected when the XFEL energy is tuned off resonances and the signals only considering resonant scattering are also shown for reference. Insets show the respective real space distribution of the resonant scattering centers $h^{res}(\vec{r})$, laser incident from left. The scattering wave vector scale bar in (a) applies to all scattering images.}
	\label{fig:results}
\end{figure*}
\begin{figure*}
	\includegraphics[width=\textwidth]{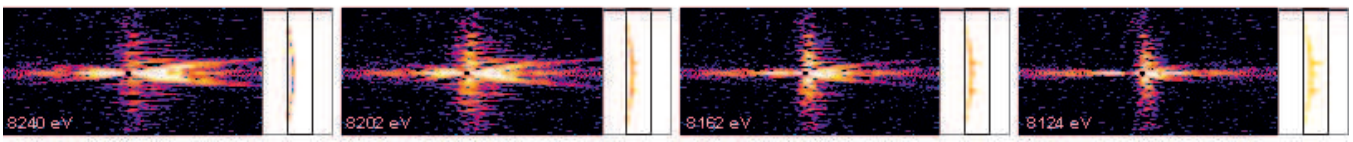}
	\caption{Scattering images for the PIC simulation employing TF ionization at four different XFEL energies corresponding to $\BoundLowE=5$ ($E_{XFEL}=8240\unit{eV}$), $\BoundLowE=6$ ($8202\unit{eV}$), $\BoundLowE=7$ ($8162\unit{eV}$)and $\BoundLowE=8$ ($8124\unit{eV}$). Insets show the respective real space distribution of the resonant scattering centers $h^{res}(\vec{r})$. All scales are the same as in Fig.~\ref{fig:results}. The scattering patterns illustrate from left to right the shift of ion distribution from compact distribution at the front to strongly filamented distribution following the electron filaments inside the foil. }
	\label{fig:results_E_scan}
\end{figure*}
\begin{figure}
	\includegraphics[width=\columnwidth]{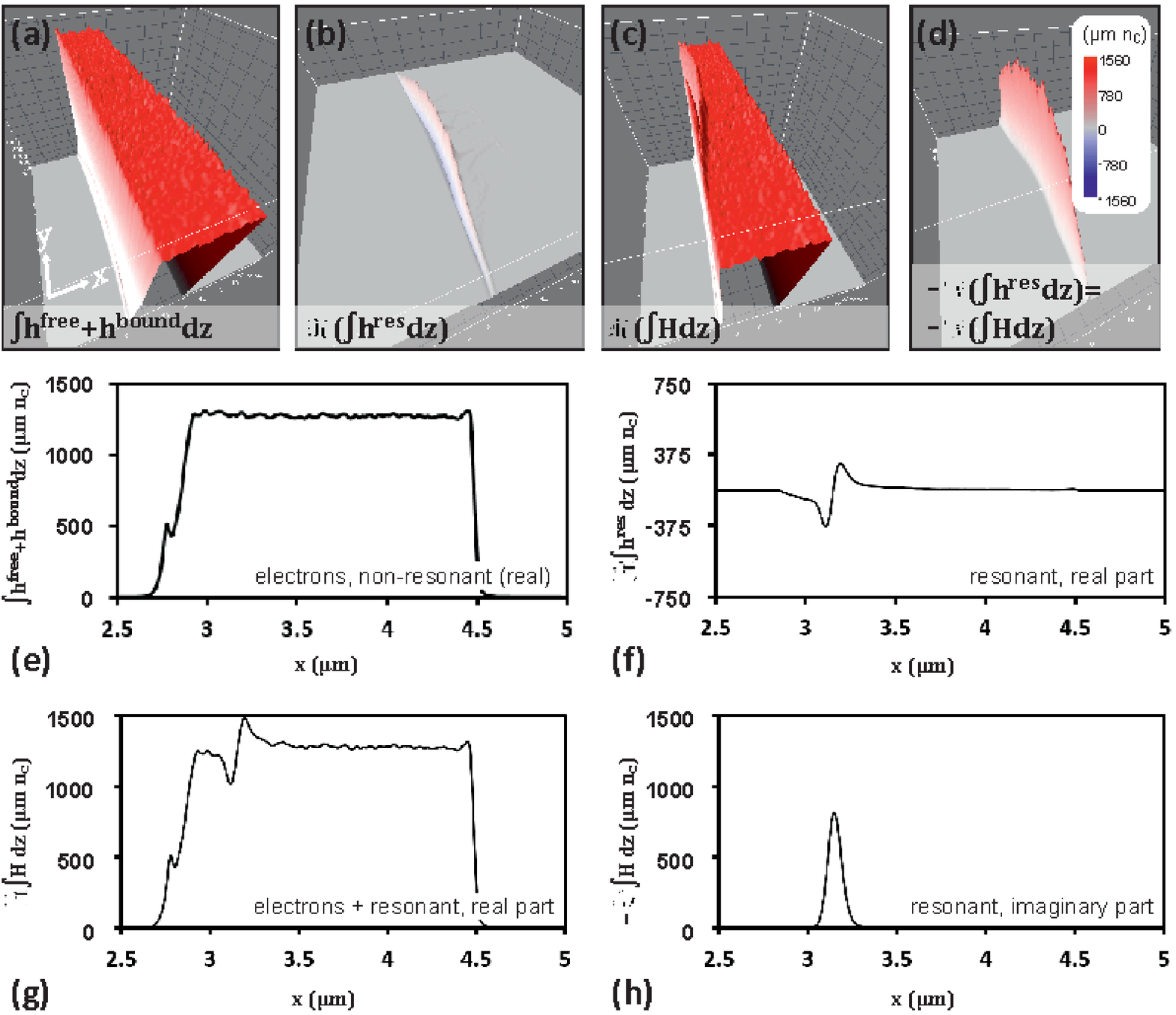}
	\caption{Real space areal density of the target model multiplied with form factor and optical corrections, $h$ for the simulation employing the TF ionization model (Fig.~\ref{fig:results}(a)). (The absolute zero was shifted in each frame since only differences are relevant for the structure of the signal.)
	 (a) $h^\mathrm{free}+h^\mathrm{bound}$, (b)
	$\Re{(h^\mathrm{res})}$, (c) $\Re(H)=h^\mathrm{free}+h^\mathrm{bound}+\Re{(h^\mathrm{res})}$ and (d) $\Im(H)=\Im{(h^\mathrm{res})}$. 
	(e-h) Respective lineouts along the laser beam axis. }
	\label{fig:A_real}
\end{figure}
\subsection{PIC simulations}
In the remainder of the paper we want to demonstrate the RCXD technique by example of simulation. 
We simulate RCXD images based on synthesized data from particle-in-cell (PIC) simulations of the laser interaction with a solid using the 2 dimensional collisional PIC code iPICLS2D~\cite{picls} including field-ionization~\cite{Landau} and collisional ionization.  
We demonstrate the feasibility, sensitivity and physical interpretation of the method by comparison of two different well established collisional ionization models, the Thomas-Fermi (TF) model~\cite{More1985} and direct impact (DI) ionization~\cite{Mishra2013}. 
In the TF model all ions are ionized up to the calculated local average charge state, thus producing a narrow distribution and neglecting the spread in charge states. 
The DI model is based on binary collisions generating a broad charge state distribution but neglects recombination and only considers ground state ionization potentials; both models neglect internal excitations. 
We approximate the local distribution of configurations ($F'_\BoundLowE$ and $F''_\BoundLowE$) at the XFEL arrival time at the target foil $t_{arr}$ during post-processing with SCFLY, computing the temperature input-parameter using the local average energy $T_e(\vec{r},t_{arr})$ from the PIC simulation with a time history proportional to that of the average over the simulation-box volume $V$, $T_e(\vec{r},t)=T_e(\vec{r},t_{arr})\left\langle T_e(\vec{r},t) \right\rangle_V/\left\langle T_e(\vec{r},t_{arr}) \right\rangle_V$. 
The most important parameters used for the simulations are given in Tab.~\ref{tab:parameters}. 

Since the PIC simulations are only 2D in space we have to use a model to extend the plasma into the third dimension (XFEL propagation direction $z$), in order to calculate the scattering signal. 
We define the target thickness in this direction to be $D_z=1\unit{\mum}$ but with only a central layer of $d_z=0.2\unit{\mum}$ thickness being resonant, while scattering in the remaining target volume only adds to the non-resonant background. 
Additionally we assume invariance of the plasma both in the resonant and non-resonant target volume in the direction of XFEL propagation. 
This setup corresponds to a sandwich target with the material of interest buried into material of neighboring Z. 
The resonant layer and its thickness is chosen to be smaller than the typical structure size. 
The 2D simulations are thus a good approximation of the real 3D situation while the non-resonant surrounding enables us to approximate the resonant vs. non-resonant signal ratios. 

Fig.~\ref{fig:results} shows the simulated scattering patterns as they would be recorded by a detector with a pixel size of $15\unit{\mum}$ at $1\unit{m}$ distance for the simulation employing the TF (a) and DI ionization model (c); Fig.~\ref{fig:results}(b) shows the scattering pattern using the TF simulation but recalculating the ion distribution during post-processing by SCFLY based on the local plasma temperature and density in order to obtain more realistic and self-consistent ion distributions and ion configurations than TF alone. \\
The lower left panels show the scattering signal taking into account only the free and bound electrons, i.e. no resonant transitions. 
This signal would be obtained far off any resonant energy and can clearly be separated in $\vec{k}$-space from the resonant part. 
Since the electron distribution $h^\mathrm{free}(\vec{r})+h^\mathrm{bound}(\vec{r})$ is rather smooth besides the sharp surface borders, the signal structure in transverse direction is weak and does not contain much useful information. \\
The resonant signals (using only $h_{res}$ in \eqref{eqn:scatter}; lower right panels) and total signals show much more structure. 
This is, in this specific example, due to the filaments in ion distribution and the ion front following the UHI laser pulse curvature, giving rise in the scattering pattern to transverse lines and a ``bow-tie'' like structure, respectively. 
Depending on the specific filament spatial frequency and width the scattering structure seen in transverse direction changes (cp. Fig.~\ref{fig:results_E_scan}), which can be used to infer the electron dynamics that imprints on ionization~\cite{lingen} as well as to benchmark ionization models. \\
Additionally, a strong asymmetry is seen in the resonant scattering signal which is due to a difference in ion distribution for different charge states. 
As shown before, ions with differing $\BoundLowE$  do not add significantly to the imaginary part of the optical correction. 
However, this is not strictly true for the real part. 
Hence, the real and imaginary parts of $H(\vec{k})$ are not only differing in amplitude (as they would if considering only one charge state and neglecting non-resonant electron scattering ), but also in phase. 
The strong asymmetry in the TF case is indicating that ions with different $\BoundLowE$ are spatially well separated while its absence in the case of DI is due to its broad charge distribution. \\
By performing a phase retrieval on the scattering pattern, the real space distribution $H(\vec{r})$ could be regained, individually for the real and imaginary part as shown in Fig.\ref{fig:A_real} for the TF case (here directly taken from PIC simulation as used to generate the scattering patterns). 
Since the non-resonant signal corresponds mainly to that of a flat homogenous foil, its scattering signal is predominantly along the direction perpendicular to the foil surfaces. 
All signal with a transverse component are hence due to resonant scattering, and as expected the optical correction's real and imaginary part differ explaining the observed asymmetry. 
The real part exhibits the expected behavior in laser direction: at the foil surface the temperature is expected to be higher due to the laser irradiation, hence there should be more ions with inner shell holes (smaller $\BoundLowE$) than inside the foil. 
With $E_0$ centered at the resonance of ions with $\BoundLowE=4$ this means more ions with $\BoundLowE=3$ and hence negative real part of the optical correction at the surface and ions with $\BoundLowE=5$ and positive real part inside the foil. 
Directly at the position of the ions with $\BoundLowE=4$, the peak in the imaginary part, the real part is zero. 

\begin{figure}
  \includegraphics[width=8cm]{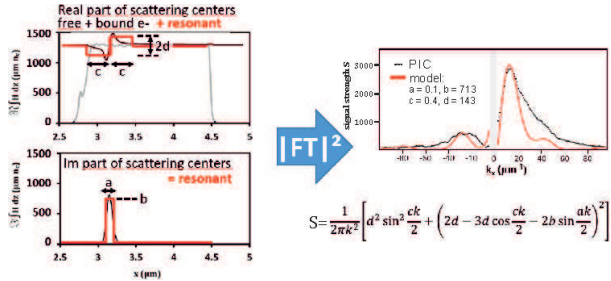}
   \caption{From the asymmetry of the scatter signal it is possible to infer the real and imaginary parts of $H(\vec{r})$ via phase retrieval. For the simulation employing TF ionization (Fig.~\ref{fig:results}(a)), already from a very simple model (red curve) approximating $\Re{(h(z))}$ and $\Im{(h(z))}$ by rectangular shapes the width a, product of form factor and density and depth b, and position c of the resonant ion front can be obtained by fitting the power spectrum (eqn. under the right hand figure) to the simulated scattering data integrated over $k_y$ (ommitting the center three lines around $k_y=0$ containing the scattering of the bulk foil).}
	\label{fig:mpr}
\end{figure}

To illustrate this we demonstrate how a simple poor-man's phase retrieval can be used to derive the position, width and density of resonant ions and ions with one electron more or less in the lower shells, as well as the local plasma temperature. 
We use the PIC data employing the TF ionization model at the resonant energy for $\BoundLowE=4$, $E_0=8290\unit{eV}$ (Fig.~\ref{fig:results}a). 
We first approximate the real- and imaginary part of the scattering centers $H(\vec{r})$ inside the target (cp. Fig.~\ref{fig:A_real}) with box functions as shown in Fig.~\ref{fig:mpr}. 
Computing the power spectrum for this model distribution we can fit the analytical result to the PIC synthesized scattering signal and obtain the density and position of resonant ions, as shown in Fig.~\ref{fig:mpr}. 

\section{Discussion and outlook}
In this paper we have presented a novel diagnostic method that will enable for the first time to measure important quantities in laser created plasmas overcoming limitations in spatial and temporal resolution of existing diagnostics. 
Tuning the XFEL energy to a specific energy selects a specific nuclear element with specific number of electrons in the inner atomic shells that will dominate the small angle scattering signal. 
Specificity depends on the XFEL energy and plasma temperature as the individual strength of resonances depends on the temperature dependent level populations but is generally possible to be better than $70\%-80\%$. 
The signals contain the full mode structure of ion-ion correlations that could be regained by numerical phase retrieval or holographic methods. 
The temperature sensitive variation of the opacity spectral shape can in turn be used to measure the spatial distribution of the instantaneous plasma temperature by using multiple-wavelength XFEL beams, and can hence be applied for equation of state measurements. 
The presented method allows for unique measurements of the ion distribution inside a warm or hot solid density plasma with high spatial and temporal resolution at the same time and hence can be used to directly test and improve ionization models on sub-micron few femtosecond scales for the first time.  
With refined simulations and theories -- including e.g. self-consistently the ionization and excitation by laser accelerated electrons, plasma expansion, temporal evolution of laser heating and cooling, recombination and de-excitation -- one can derive the distribution of ion electronic configurations and hence construct the local history of ionization, electron energy distribution and density and eventually determine effects of non-thermal and transient electron dynamics on ionization dynamics. 
The spatial and temporal resolution are only limited by the XFEL wavelength and pulse duration. 
The temporal evolution of the properties may even be studied on femtosecond time scales employing more advanced schemes such as photon correlation spectroscopy using split and delayed pulses~\cite{Grubel2007,Kluge2014}. \\
The presented method allows for the first time space and time resolved dynamical (via pump-probe or correlation spectroscopy) measurement on a few-nanometer, few-femtosecond level of many important properties of high-power laser generated plasmas such as the hole-boring, shock front and ionization front position, width and speed, mode structure of ion filaments and their respective growth rates and plasma temperature, potentially opening up a whole new field for understanding transient laser-plasma dynamics on a level not accessible to experiments before. 
\begin{acknowledgments}
The work has been partially supported the German Federal Ministry of Education and Research (BMBF) under
contract number 03Z1O511. The authors thank J. Grenzer, R. Pausch, K. Steiniger (HZDR) and Y. Sentoku (U Reno) for fruitful discussions. 
\end{acknowledgments}

%

\end{document}